\documentclass{article}
\usepackage{amsmath}
\usepackage{amsfonts}
\usepackage{amssymb}
\usepackage{graphicx}
\begin{document}

\title{Analysis of the self-similar spreading of power law fluids}
\author{D. G. Aronson\thanks{Mathematics Department, University of Minnesota, 127
Vincent Hall, 206 Church street, Minneapolis, MN-55444, USA.}, S. I.
Betel\'{u}\thanks{Mathematics Department, University of North Texas, P.O. Box
311430, Denton, TX 76203-1430, USA.}, M. A. Fontelos, A. S\'{a}%
nchez\thanks{Departamento de Matem\'{a}tica Aplicada, Universidad Rey Juan
Carlos, C/ Tulip\'{a}n S/N, M\'{o}stoles 28933, Madrid, Spain.}}
\maketitle
\begin{abstract}
We consider the equation that models the spreading of thin liquid films of
power-law rheology. In particular, we analyze the existence and uniqueness of
source-type self-similar solutions in planar and circular symmetries. We find
that for shear-thinning fluids there exist a family of such solutions representing both
finite and zero contact angle drops and that the solutions with zero contact
angle are unique. We also prove the existence of traveling waves in one space dimension
and classify them.
\end{abstract}

\section{Introduction}

Here we study capillary spreadings of thin films of liquids of power-law
rheology, also known as Ostwald-de Waele fluids \cite{Bird}. The power-law
rheology is one of the simplest generalizations of the Newtonian one, in which
the effective viscosity $\eta$ at a point is assumed to be a function of the
local rate of deformation $\dot{\gamma}$ given by $\eta=m|\dot{\gamma}|^{1/\lambda-1}$.
 The values of $m$ and $\lambda$ depend on the physical
properties of the liquid. When $\lambda>1$ the viscosity tends to zero at high
strain rates \cite{Bird} and is larger at low strain rates (these fluids are
called shear-thinning).

In \cite{King,Betelu} the following equation for one dimensional motion was
derived using the lubrication approximation:
\begin{equation}
u_{t}+(u^{\lambda+2}|u_{xxx}|^{\lambda-1}u_{xxx})_{x}=0.\label{ecu}%
\end{equation}
where $u(x,t)$ represents the thickness of the one-dimensional liquid film at
position $x$ and time $t$. In \cite{Ansini} a
generalized version of (\ref{ecu}) was studied by means of asymptotic and
perturbative techniques in order to construct approximate solutions
representing the spreading of a droplet. We will look for solutions with a compact support
$\left[  x_{1}(t),x_{2}(t)\right]  $. Therefore, by conservation of mass and
equation (\ref{ecu}),
\[
0=\frac{d}{dt}\int_{x_{1}(t)}^{x_{2}(t)}u(x,t)dx
\]%
\[
=x_{2}^{\prime}(t)u(x_{2}(t),t)-x_{1}^{\prime}(t)u(x_{1}(t),t)+\left.  \left(
u^{\lambda+2}|u_{xxx}|^{\lambda-1}u_{xxx}\right)  \right|  _{x_{1}(t)}%
^{x_{2}(t)}%
\]
which forces us to impose
\begin{equation}
x_{i}^{\prime}(t)u(x_{i}(t),t)+\left(  u^{\lambda+2}|u_{xxx}|^{\lambda
-1}u_{xxx}\right)  (x_{i}(t),t)=0\;,\;i=1,2\text{.}\label{noflux}%
\end{equation}

Equation (\ref{ecu}) formally admits solutions of the form
\begin{equation}
u(x,t)=\frac{A}{t^{\beta}}U\left(  \frac{x}{t^{\beta}}\right) \label{sol}%
\end{equation}
with
\begin{equation}
\beta=\frac{1}{5\lambda+2}.
\end{equation}
Without loss of generality, one can set $A^{2\lambda+1}=1/(5\lambda+2)$. Then with $\eta=x/t^\beta$, 
$U(\eta)$ satisfies the ODE,
\begin{equation}
U^{\lambda+2}{U^{\prime\prime\prime}}^{\lambda}=\eta U\;,\label{selfsimilar}%
\end{equation}
which results from introducing (\ref{sol}) into (\ref{ecu}), integrating once
and choosing the integration constant $K=0$ (by(\ref{noflux})).

The initial conditions for symmetric drops are
\begin{equation}
U(0)=1,\hspace{1cm}U^{\prime}(0)=0,\hspace{1cm}U^{\prime\prime}(0)=-\kappa
,\label{algo}%
\end{equation}
where $\kappa$ is a real positive parameter.

For radially symmetric flows, the equation of motion is (cf. \cite{BF}):
\[
u_{t}+\frac1r\left(  ru^{\lambda+2}\left|  \left(  u_{rr}+\frac{u_{r}%
}r\right)  _{r}\right|  ^{\lambda-1}\left(  u_{rr}+\frac{u_{r}}r\right)
_{r}\right)  _{r}=0,
\]
where $r$ is the radial coordinate, and the self-similar solutions are of the form
\[
u(r,t)=\frac A{t^{2\beta}}U\left(  \frac r{t^{\beta}}\right)
\]
with
\[
\beta=\frac1{7\lambda+3}%
\]
and $U(\eta)$ satisfying the following ordinary differential equation:
\begin{equation}
U^{\lambda+2}\left(  \left(  U^{\prime\prime}+\frac{U^{\prime}}\eta\right)
^{\prime}\right)  ^{\lambda}=\eta U\;.\label{radial1}%
\end{equation}
Analogously to (\ref{algo}) we impose
\begin{equation}
U(0)=1,\hspace{1cm}U^{\prime}(0)=0,\hspace{1cm}\lim_{\eta\rightarrow0^{+}%
}\frac1\eta(\eta U^{\prime})^{\prime}=-\kappa\;.\label{radial2}%
\end{equation}

For simplicity and clarity we begin with a datached analysis of the existence 
and uniqueness of one-dimensional self-similar solutions, i.e., solutions to 
Eqs. (\ref{selfsimilar}-\ref{algo}). This is done in section 2. In section 3 we indicate the
 changes needed to deal with radially symmetric self-similar solutions
 (Eqs. (\ref{radial1}-\ref{radial2})). Finally, in an Appendix we study the traveling wave
solutions of (\ref{ecu}) and classify all the possible behaviors of moving
fronts close to the interface.

\section{Analysis of one dimensional self-similar solutions}

In order to remove the parameter $\kappa$ from the initial condition we introduce the change
of variables
\begin{equation}
x= \eta\sqrt{\kappa}.
\end{equation}
Then $z(x,\gamma)=U(\eta)$ satisfies
\begin{align}
z^{1+a}z^{\prime\prime\prime} =\gamma x^{a}\\
z(0)=1,z^{\prime}(0)=0,z^{\prime\prime}(0)=-1,\nonumber
\end{align}
with
\begin{equation}
\gamma= k^{-(3+1/\lambda)/2 }%
\end{equation}
and
\[
a=1/\lambda<1.
\]
In the following subsections we prove:\\

\textbf{Theorem 1} For each $a\in(0,1)$ there exists a $\gamma=\gamma(a)$ and
$y=y(a)$ such that $z(x,\gamma(a))$ first reaches $z=0$ for $x=y(a)$ and
$z^{\prime}(y(a),\gamma(a))=0$. The function $z(x,\gamma(a))$ satisfying these
conditions is unique.

\textbf{Theorem 2} Given $a\in(0,1)$ and $\theta\in(0,1)$ there exist
$\gamma=\gamma_{\theta}$ and $y=y_{\theta}$ such that $z(x,\gamma_{\theta})$
first reaches $z=0$ for $x=y_{\theta}$ and $z^{\prime}(y_{\theta}%
,\gamma_{\theta})=-\sqrt{2}\theta$.\newline 

These results are physically meaningful, because they imply that for
$\lambda>1$ there exist solutions describing one-dimensional drops with fronts
advancing at finite speed. This result contrasts with the Newtonian case
$\lambda=1$, where such solutions do not exist \cite{Bernis}.

We also show that source-type self-similar solutions cannot exist for
$\lambda\le1$.

All these results were suggested by numerical calculations in \cite{Betelu}
and asymptotic analysis in \cite{Betelu,King}.

In an appendix we will show, by studying traveling wave solutions, that the
local behaviors of moving fronts near the interface are just those described
in Theorems 1 and 2.

\subsection{\bigskip\ General Properties of Solutions.}

We consider the initial value problem
\begin{equation}
z^{1+a}z^{\prime\prime\prime}=\gamma x^{a}\label{a}
\end{equation}%
\[
z(0)=1,z^{\prime}(0)=0,z^{\prime\prime}(0)=-1,
\]
where $a\geq0$ is fixed and $\gamma\in\mathbf{R}$ is a parameter. We are
interested in finding those values of $\gamma$ for which there are interfaces,
i.e., such that the solution $z=z(x,\gamma)$ has a zero for a finite value of
$x$ and $\gamma$.

If $\gamma=0$ then the non-trivial solution to (\ref{a}) is
\begin{equation}
z(x,0)=1-\frac{x^{2}}{2}\label{b}
\end{equation}
which decreases from $1$ to $0$ as $x$ increases from $0$ to $\sqrt{2}$. The
interface is $x=\sqrt{2}$.

If $\gamma<0$ then the first three $x$-derivatives of $z(x,\gamma)$ are
negative whenever $z>0$. Thus $z$ is a decreasing function of $x$ as long as
$z$ is positive. Moreover, $z^{\prime\prime\prime}<0$ and the initial
conditions imply that $z(x,\gamma)<z(x,0)$ for $x>0$. Physically, the case
$\gamma<0$ represents retracting, dewetting drops.

If $\gamma>0$ then $z(x,\gamma)>z(x,0)$ for $x>0.$ Moreover, $z^{\prime
}(x,\gamma)<0$ at least for all sufficiently small $x>0$. Either $z^{\prime
}(x,\gamma)<0$ for all $x$ such that $z(x,\gamma)>0$ or else there exists a
$\zeta>0$ such that $z(x,\gamma)$ decreases on $(0,\zeta)$ and achieves a
minimum at $x=\zeta$ with $z(x$,$\gamma)\in(0,1)$. At $x=\zeta$ we have
$z^{\prime}(\zeta,\gamma)=0$ and $z^{\prime\prime}(\zeta,\gamma)\geq0$. Since
$z^{\prime\prime\prime}>0$ it follows that $z^{\prime\prime}(x,\gamma)>0$ and
$z^{\prime}(x,\gamma)>0$ for all $x>\zeta$. Therefore if $z$ is not everywhere
decreasing for $z>0$ then $z$ has a unique minimum at some $x=\zeta>0$ with
\[
0<z(\zeta,\gamma)<z(x,\gamma)<1,\text{ }z^{\prime}(x,\gamma)<0\text{ for }%
x\in(0,\zeta)
\]
and
\[
z(\zeta,\gamma)<z(x,\gamma),\text{ }z^{\prime}(x,\gamma)>0\text{ for \ }%
x\in(\zeta,\infty).
\]

\subsection{Local Expansions Near the Interface}

There are two different asymptotic expansions for the solution near the root
$z(y,\gamma)=0$: one represents solutions with finite contact angle,
\begin{equation}
z(x,\gamma)=\sqrt{2}\theta(y-x)+\frac{\gamma y^{a}}{2^{\frac{1+a}2}%
\theta^{1+a}a(1-a)(2-a)}(y-x)^{2-a}+\mbox{lower order terms}\label{exp1}%
\end{equation}
where $\theta\in\left(  0,1\right)  $, and another with zero contact angle
\begin{equation}
z(x,\gamma)=\left(  \frac{\gamma(2+a)^{3}}{3a(1-a)(1+2a)}\right)
^{1/(2+a)}y^{a/(2+a)}(y-x)^{3/(2+a)}+\mbox{lower order
terms}.\label{exp2}%
\end{equation}

It is simple to show the nonexistence of local expansions near the interface
if $a\geq1$ and $\gamma>0$. If $a\geq1$ and $\gamma\leq0$, then the local
expansion is given by (\ref{exp1}).

We can write formal representation formulae for the solution to (\ref{a}). Let
\[
I_{1}(x,\gamma)=\int_{0}^{x}\frac{\xi^{a}d\xi}{z^{1+a}(\xi,\gamma)}%
\]
and
\[
I_{j+1}(x,\gamma)=\int_{0}^{x}I_{j}(\zeta,\gamma)d\zeta
\]
for $j=1,2$. Then as long as $z>0$%
\begin{equation}
z^{\prime\prime}(x,\gamma)=-1+\gamma I_{1}(x,\gamma),\label{c.i}
\end{equation}%
\begin{equation}
z^{\prime}(x,\gamma)=-x+\gamma I_{2}(x,\gamma),\label{c.ii}
\end{equation}
and
\begin{equation}
z(x,\gamma)=1-\frac{x^{2}}{2}+\gamma I_{3}(x,\gamma).\label{c.iii}
\end{equation}

It is clear from (\ref{exp1}) and (\ref{exp2}) that if $a<1$ then the
representation formulae (\ref{c.ii}) and (\ref{c.iii}) can be extended to the interface.
In particular, if $z(x,\gamma)>0$ for $0\leq x<y$ and $z(x,\gamma
)\rightarrow0$ as $x\nearrow y$ then
\begin{equation}
z^{\prime\prime}(x,\gamma)\rightarrow+\infty,\label{d.i}
\end{equation}
\begin{equation}
z^{\prime}(x,\gamma)\rightarrow z^{\prime}(y,\gamma)=-y+\gamma I_{2}
(y,\gamma),\label{d.ii}
\end{equation}
and
\begin{equation}
z(x,\gamma)\rightarrow z(y,\gamma)=1-\frac{y^{2}}{2}+\gamma I_{3}%
(y,\gamma).\label{d.iii}
\end{equation}
If $a\geq1$ then (\ref{d.i}) and (\ref{d.iii}) are still valid, but (\ref{d.ii}) is replaced by
$z^{\prime}(x,\gamma)\rightarrow+\infty$. Hence, there are no interfaces for
$a\geq1$ unless $\gamma\leq0$.

There are no interfaces for sufficiently large values of $\gamma$. More
precisely, we prove

\textbf{Lemma 1.}\textit{\ There are no solutions with an interface if}
\begin{equation}
\gamma>\left\{  \frac{1+a}{2(3+a)}\right\}  ^{\frac{1+a}{2}}(1+a)(2+a).\label{e}
\end{equation}

\textbf{Proof.} As we observed above, any positive local minimum of
$z(x,\gamma)$ is in fact a unique global minimum. Thus if $z$ has a positive
minimum, then $z$ does not have an interface. Suppose $z$ has a positive
minimum at $x=x_{0}$. Then since $0<z<1$, we have
\[
z(x,\gamma)=1-\frac{x^{2}}{2}+\gamma I_{3}(x,\gamma)\geq1-\frac{x^{2}}%
{2}+\frac{\gamma x^{3+a}}{(1+a)(2+a)(3+a)}\equiv P(x,\gamma)
\]
for $x\in\lbrack0,x_{0}]$. The condition (\ref{e}) guarantees that the positivity of
the minimum value of $P$. Therefore if (\ref{e}) is satisfied, $z$ is everywhere
positive and has no interface.$\square$

\bigskip

\subsection{Analysis for Small $\gamma$}

Since (\ref{a}) is singular at $z=0$, we shall first study the behavior of the
solutions for $z>0$, and, instead of studying the interfaces $z(y,\gamma)=0$
directly, we first solve
\begin{equation}
z(y,\gamma)=\delta\label{f}
\end{equation}
for given $\delta\in(0,1)$. In the next subsection we study the limit as
$\delta\searrow0$. We begin by studying (a) in the neighborhood of the
solution (\ref{b}) for small $\left|  \gamma\right|  $.

\bigskip We have
\[
z(y_{1}(\delta),0)=\delta\text{ and }z^{\prime}(y_{1}(\delta),0)=-y_{1}
(\delta)<0,
\]
where $y_{1}(\delta)=\sqrt{2(1-\delta)}$. Since $z(x,\gamma)$ is smooth for
$z>0$, by the Implicit Function Theorem there exists a smooth function
$y(\gamma,\delta)$ for sufficiently small $\left|  \gamma\right|  $ such that
\[
y(0,\delta)=y_{1}(\delta)\text{ and }z(y(\gamma,\delta),\gamma)=\delta.
\]
$y(\gamma,\delta)$ exists for all $\gamma$ such that
\begin{equation}
z^{\prime}(y(\gamma,\delta),\gamma)<0.\label{g}
\end{equation}
As we observed above, (\ref{g}) holds for all $\gamma\leq0$. Moreover, since (\ref{g})
holds for $\gamma=0,$ and since $y(\gamma,\delta)$ depends smoothly on
$\gamma$ and \ $z\left(  y,\gamma\right)  $ depends smoothly on $y$ it follows
that (\ref{g}) continues to hold for all sufficiently small $\gamma>0$.
Differentiating (\ref{f}) with respect to $\gamma$ we find
\[
z^{\prime}(y,\gamma)\frac{\partial y}{\partial\gamma}+z_{\gamma}(y,\gamma)=0.
\]
Therefore
\[
\frac{\partial y}{\partial\gamma}\mid_{\gamma=0}=-\frac{z_{\gamma}%
(y_{1}(\delta),0)}{z^{\prime}(y_{1}(\delta),0)}=\frac{I_{2}(y_{1}(\delta
),0)}{\sqrt{2(1-\delta)}}>0
\]
so that $y$ is an increasing function of $\gamma$ at least for $\left|
\gamma\right|  $ small.

\bigskip

\subsection{Estimates for solutions with interfaces}

\bigskip

We now investigate the maximal $\gamma$-interval of existence for the
functions $y(\gamma,\delta)$ constructed above. In the limit as $\delta
\rightarrow0$ this will lead to solutions with zero contact angle.

In view of (\ref{g}) we define
\[
\gamma_{0}(\delta)\equiv\sup\left\{  \gamma>0:z^{\prime}(y(\gamma
,\delta),\gamma)<0\right\}
\]
and
\[
y_{0}(\delta)\equiv\lim_{\gamma\nearrow\gamma_{0}(\delta)}y(\gamma,\delta).
\]
Note that since $z^{\prime}(y,\gamma)$ and $y(\gamma,\delta)$ are continuous
we have
\[
z(y_{0}(\delta),\gamma_{0}(\delta))=\delta\text{ and }z^{\prime}(y_{0}%
(\delta),\gamma_{0}(\delta))=0.
\]
We now derive some estimates for $\gamma_{0}(\delta)$ and $y_{0}(\delta)$.

\bigskip\textbf{Lemma 2.}\textit{\ Let}
\[
B(\gamma)=\left\{  \frac{(1+a)(2+a)}{\gamma}\right\}  ^{\frac{1}{1+a}%
}\text{\textit{and} }G=2^{\frac{1+a}{2}}(1+a)(2+a).
\]
\textit{Then}
\begin{equation}
0<\gamma_{0}(\delta)<G\label{i}%
\end{equation}
\textit{and}
\begin{equation}
\sqrt{2(1-\delta)}<y_{0}(\delta)<B(\gamma_{0}(\delta)).\label{j}%
\end{equation}

\textbf{Proof.} For $0<x<y(\gamma,\delta)$ we have $\delta<z(x,\gamma)<1$, and
$z^{\prime}(y(\gamma,\delta),\delta)<0$. It follows from (\ref{d.ii}) that
\[
0>z^{\prime}(y(\gamma,\delta),\gamma)=-y(\gamma,\delta)+\gamma I_{2}%
(y(\gamma,\delta),\gamma)>-y(\gamma,\delta)+\frac{\gamma y^{1+a}(\gamma
,\delta)}{(1+a)(2+a)}.
\]
Therefore
\[
\sqrt{2(1-\delta)}<y(\gamma,\delta)<B(\gamma).
\]
Letting $\gamma\nearrow\gamma_{0}(\delta)$ we immediately obtain (\ref{j}). Then (\ref{i})
follows from $\sqrt{2(1-\delta)}<B(\gamma_{0}(\delta)).\square$

For arbitrary fixed $\theta\in(0,1)$ consider
\[
\Delta(\gamma,\theta)\equiv z^{\prime}(y(\gamma,\delta),\gamma)+\theta
\sqrt{2(1-\delta)}.
\]
$\Delta(\gamma,\theta)$ is a continuous function of $\gamma\in\lbrack
0,\gamma_{0}(\delta)]$ with
\[
\Delta(0,\theta)=-(1-\theta)\sqrt{2(1-\delta)}<0\text{ and }\Delta(\gamma
_{0}(\delta),\theta)=\theta\sqrt{2(1-\delta)}>0.
\]
By the Intermediate Value Property of continuous functions there exists
\ $\gamma(\delta,\theta)\in(0,\gamma_{0}(\delta))$ such that $\Delta
($\ $\gamma(\delta,\theta),\theta)=0$, i.e.,
\begin{equation}
z^{\prime}(Y(\delta,\theta),\gamma(\delta,\theta))=-\theta\sqrt{2(1-\delta
)},\label{k}%
\end{equation}
where
\[
Y(\delta,\theta)=y(\gamma(\delta,\theta),\delta).
\]
We take \ $\gamma(\delta,\theta)$ to be the smallest positive value of
$\gamma$ for which (\ref{k}) holds. We extend the definition of $\gamma
(\delta,\theta)$ by setting $\gamma(\delta,0)=\gamma_{0}(\delta)$ and
$\gamma(\delta,1)=0.$ In view of (\ref{i})
\[
\gamma_{\theta}\equiv\lim_{\delta\searrow0}\inf\gamma(\delta,\theta)
\]
exists and satisfies
\[
0\leq\gamma_{\theta}\leq G
\]
for arbitrary $\theta\in\lbrack0,1].$ For each $\theta\in\lbrack0,1]$ fix a
sequence $\{\delta_{\theta}^{j}\}$ such that $\delta_{\theta}^{j}\searrow0$
and $\gamma_{\theta}^{j}\equiv\gamma(\delta_{\theta}^{j},\theta)\rightarrow
\gamma_{\theta}$ as $j\rightarrow\infty$.

\textbf{Lemma 3.} $\gamma_{\theta}>0$ if and only if $a<1.$

\textbf{Proof. }From (\ref{c.ii}) we have
\[
-\theta\sqrt{2(1-\delta)}=z^{\prime}(Y(\delta,\theta),\gamma(\delta
,\theta))=-Y(\delta,\theta)+\gamma(\delta,\theta)I(\delta,\theta)
\]
or
\[
Y(\delta,\theta)-\theta\sqrt{2(1-\delta)}=\gamma(\delta,\theta)I(\delta
,\theta),
\]
where
\[
I(\delta,\theta)=I_{2}(Y(\delta,\theta),\gamma(\delta,\theta)).
\]
Suppose that $\gamma_{\theta}=0$. Then since
\[
(1-\theta)\sqrt{2(1-\delta_{\theta}^{j})}<\gamma_{\theta}^{j}I(\delta_{\theta
}^{j},\theta),
\]
$\gamma(\delta_{\theta}^{j},\theta)\rightarrow0$ implies $I(\delta_{\theta
}^{j},\theta)\rightarrow\infty$. In view of the asymptotic behavior of $z$
near $0$ this can only occur if $a\geq1$. On the other hand, if $\gamma
(\delta_{\theta}^{j},\theta)>0$ then
\[
\gamma(\delta,\theta)I(\delta,\theta)<Y(\delta,\theta)<B(\gamma(\delta
,\theta))
\]
so that
\[
I(\delta_{\theta}^{j},\theta)<\frac{B(\gamma_{\theta}^{j})}{\gamma_{\theta
}^{j}}.
\]
Therefore
\[
0<\lim_{j\rightarrow\infty}\sup I(\delta_{\theta}^{j},\theta)\leq
\frac{B(\gamma_{\theta})}{\gamma_{\theta}}<\infty
\]
which can occur only if $a<1.\square$

We now investigate the behavior of $Y(\delta_{\theta}^{j},\theta)$ as
$j\rightarrow\infty.$

\textbf{Lemma 4.}\textit{\ If }$a<1$\textit{\ then}
\[
\sqrt{2}\leq y_{\theta}\equiv\lim_{j\rightarrow\infty}Y(\delta_{\theta}%
^{j},\theta)\leq B(\gamma_{\theta}).
\]

\textbf{Proof.} $\gamma_{\theta}^{j}\rightarrow\gamma_{\theta}$ implies that
for arbitrary $\varepsilon\in(0,\gamma_{\theta})$ there exists an
$N=N(\varepsilon)>0$ such that $\gamma_{\theta}^{j}>\gamma_{\theta
}-\varepsilon$ for all $j>N$. In view of (\ref{j})
\[
\sqrt{2(1-\delta_{\theta}^{j})}<Y(\delta_{\theta}^{j},\theta)\leq
B(\gamma_{\theta}-\varepsilon).
\]
Thus there is a subsequence of the $\delta_{\theta}^{j}$ which we again call
$\{\delta_{\theta}^{j}\}$ along which $Y(\delta_{\theta}^{j},\theta)$
converges to a limit $y_{\theta}$ which satisfies
\[
\sqrt{2}\leq y_{\theta}\leq B(\gamma_{\theta}-\varepsilon).
\]
Since $\varepsilon$ is arbitrary the assertion now follows by letting
$\varepsilon\searrow0$.$\square$

Finally we prove

\textbf{Lemma 5.}\textit{\ If }$a<1$\textit{\ then}
\[
z(y_{\theta},\gamma_{\theta})=\lim_{j\rightarrow\infty}z(Y(\delta_{\theta}%
^{j},\theta),\gamma_{\theta})=0\text{ and }z^{\prime}(y_{\theta}%
,\gamma_{\theta})=\lim_{j\rightarrow\infty}z^{\prime}(Y(\delta_{\theta}%
^{j},\theta),\gamma_{\theta})=-\theta\sqrt{2}.
\]

\textbf{Proof. }From the representation formulas (\ref{c.ii}) we have
\[
z(Y(\delta_{\theta}^{j},\theta),\gamma_{\theta})=1-\frac{Y^{2}(\delta_{\theta
}^{j},\theta)}{2}+\gamma_{0}I_{3}(Y(\delta_{\theta}^{j},\theta),\gamma
_{\theta})
\]
which we rewrite in the form
\begin{align*}
z(Y(\delta_{\theta}^{j},\theta),\gamma_{\theta})  & =\delta_{\theta}%
^{j}+\gamma_{\theta}\left\{  I_{3}(Y(\delta_{\theta}^{j},\theta),\gamma
_{\theta})-I_{3}(Y(\delta_{\theta}^{j},\theta),\gamma_{\theta}^{j})\right\} \\
& +\left\{  \gamma_{\theta}-\gamma_{\theta}^{j}\right\}  I_{3}(Y(\delta
_{\theta}^{j},\theta),\gamma_{\theta}^{j}).
\end{align*}
The first and third terms on the right hand side clearly tend to zero as
$j\rightarrow\infty.$ The convergence of the middle term to zero follows from
the asymptotic behavior of $z$ near zero and the Lebesgue Dominated
Convergence Theorem. The result for $z^{\prime}$ is proved in a similar
manner.$\square$

\textbf{Corollary.}\textit{\ If }$a<1$\textit{\ then }$y_{\theta}>\sqrt{2}.$

\textbf{Proof.} Since $\gamma_{\theta}>0$ we have $z^{\prime\prime\prime
}(x,\gamma_{\theta})>0$ on $(0,y_{\theta}).$ Therefore
\[
0=z(y_{\theta},\gamma_{\theta})>1-\frac{y_{\theta}^{2}}{2}%
\]
which implies that $y_{\theta}>\sqrt{2}.\square$

\bigskip The analytic considerations given above establish the existence for
$a<1$ of solutions whose interfaces have zero contact angle and of solutions
whose interfaces have non-zero contact angles. In particular, for each
$\theta\in[0,1]$ there is a pair $(y_{\theta},\gamma_{\theta})$ such that the
solution $z(x,\gamma_{\theta})$ has its interface at $x=y_{\theta}$ with
$z^{\prime}(y_{\theta},\gamma_{\theta})=$ $-\theta\sqrt{2}$. However, the
results so far do not give any global information about uniqueness and
monotonicity of $(y_{\theta},\gamma_{\theta})$. In the next subsection we will
prove the uniqueness of zero contact angle solutions for each $\gamma>0$.

\subsection{Uniqueness of zero contact angle solutions}

\bigskip In this section we present a proof of uniqueness of the zero contact
angle solution following arguments similar to those used in \cite{BHK} in a
somewhat different context. Rewrite equation (\ref{selfsimilar}) in the form
\begin{equation}
U^{\prime\prime\prime}=\eta^{\frac{1}{\lambda}}U^{-s},\label{ec1}%
\end{equation}
where $s=1+1/\lambda$, and \ let $U_{j}(\eta)$ for $j=1,2$ be solutions to
(\ref{ec1}) such that
\[
U_{j}(0)=1,U_{j}^{\prime}(0)=0,
\]
and
\[
U_{j}(\eta_{j})=U_{j}^{\prime}(\eta_{j})=0.
\]
Without loss of generality we assume that
\[
\eta_{1}\geq\eta_{2}.
\]

Set
\[
c=\left(  \frac{\eta_{1}}{\eta_{2}}\right)  ^{\frac{3\lambda+1}{2\lambda+1}%
}\text{ and }\xi=\frac{\eta\eta_{1}}{\eta_{2}}.
\]
Note that $c\geq1$. The rescaled function
\[
\widetilde{U}_{2}(\xi)=cU_{2}(\eta)
\]
is a solution to (11) with
\[
\widetilde{U}_{2}(0)=c,\widetilde{U}_{2}^{\prime}(0)=0,
\]
and
\[
\widetilde{U}_{2}(\eta_{1})=\widetilde{U}_{2}^{\prime}(\eta_{1}^{-})=0.
\]

For $\xi\in\lbrack0,\eta_{1}]$ define
\[
V(\xi)\equiv\widetilde{U}_{2}(\xi)-U_{1}(\xi).
\]
Then
\[
V(0)=c-1\geq0,V^{\prime}(0)=0,\text{ and }V(\eta_{1})=0.
\]
Moreover, in view of (\ref{ec1}),
\[
V^{\prime\prime\prime}=\xi^{\frac{1}{\lambda}}\left\{  f(\widetilde{U}%
_{2})-f(U_{1})\right\}  ,
\]
where
\[
f(\tau)=\tau^{-s}.
\]
Note that $f$ is a monotone decreasing function with $f^{\prime}<0$ for
$\tau>0.$ By the Mean Value Theorem
\[
VV^{\prime\prime\prime}=\xi^{\frac{1}{\lambda}}f^{\prime}(W)V^{2}%
\]
for some $W$ between $\widetilde{U}_{2}$ and $U_{1}.$ Thus
\begin{equation}
VV^{\prime\prime\prime}\leq0\text{ on }[0,\eta_{1})\text{.}\label{ineq}%
\end{equation}

Set
\[
h(\xi)=V(\xi)V^{\prime\prime}(\xi)-\frac{1}{2}V^{\prime2}(\xi).
\]
Then
\[
h(0)=(c-1)V^{\prime\prime}(0)\text{ and }h^{\prime}=VV^{\prime\prime\prime
}\leq0\text{ on }[0,\eta_{1}),
\]
i.e., $h$ is a monotone decreasing function on $[0,\eta_{1}]$. Using the
asymptotic expansion (\ref{exp2}) we find that for sufficiently large
$\xi<\eta_{1}$%
\[
h(\xi)\sim K(\eta_{1}-\xi)^{\frac{2(1-a)}{2+a}}%
\]
for some constant $K$. Since $a<1$ it follows that
\[
h(\eta_{1}^{-})=0.
\]

If $h(0)\leq0$ then $h$ decreasing and $h(\eta_{1}^{-})=0$ imply that
$h\equiv0$ on $[0,\eta_{1}]$. Thus
\[
h^{\prime}=\xi^{\frac{1}{\lambda}}(\widetilde{U}_{2}^{-s}-U_{1}^{-s}%
)(\widetilde{U}_{2}-U_{1})=0\text{ on }[0,\eta_{1})
\]
and this implies that $\widetilde{U}_{2}=U_{1}$ on $[0,\eta_{1})$. In
particular, $c=\widetilde{U}_{2}(0)=U_{1}(0)=1$. It follows that $\eta
_{1}=\eta_{2}$ so that
\[
U_{2}(\xi)=\widetilde{U}_{2}(\xi)=U_{1}(\xi)\text{ on }[0,\eta_{1}].
\]

Suppose that $h(0)>0.$ Together with $h(\eta_{1}^{-})=0$ and the fact that $h$
is decreasing this implies that $h\geq0$ on $[0,\eta_{1}]$. Therefore
\begin{equation}
VV^{\prime\prime}\geq\frac{1}{2}V^{\prime2}\geq0\text{ on }[0,\eta
_{1}].\label{ec3}%
\end{equation}
Moreover, it follows from $h(0)>0$ that $c>1$ and $V^{\prime\prime}(0)>0.$
Hence there exists a $\zeta_{1}\in(0,\eta_{1})$ such that
\[
V\geq c-1>0\text{ on }[0,\zeta_{1}]\text{ and }V^{\prime}(\zeta_{1})>0.
\]
On the other hand, $V(\eta_{1})=0$ means that $V$, which is initially
increasing, must start to decrease somewhere in $(\zeta_{1},\eta_{1})$. Thus
there exists a $\zeta_{2}\in(\zeta_{1},\eta_{1})$ such that
\[
V>0\text{ on }[0,\zeta_{2}]\text{ and }V^{\prime}(\zeta_{2})<0.
\]
It follows that there exists a $\zeta_{3}\in(\zeta_{1},\zeta_{2})$ such that
\[
V(\zeta_{3})V^{\prime\prime}(\zeta_{3})<0
\]
and this contradicts (\ref{ec3}). We conclude that we must have $h(0)\leq0$
and, as we have shown above, this implies that $U_{1}\equiv U_{2}.$

\section{The radially symmetric case}

In this section we will briefly describe how the proof of existence and
uniqueness for the 1-D model extends to radially symmetric solutions. In this
case, the evolution equation is (cf. \cite{BF}):
\[
u_{t}+\frac{1}{r}\left(  ru^{\lambda+2}\left|  \left(  u_{rr}+\frac{u_{r}}%
{r}\right)  _{r}\right|  ^{\lambda-1}\left(  u_{rr}+\frac{u_{r}}{r}\right)
_{r}\right)  _{r}=0
\]
and the self-similar solutions are of the form
\[
u(r,t)=\frac{A}{t^{2\beta}}U\left(  \frac{r}{t^{\beta}}\right)
\]
with
\[
\beta=\frac{1}{7\lambda+3}%
\]
and $U(\eta)$ satisfying the following ordinary differential equation:
\begin{equation}
U^{\lambda+2}\left(  \left(  U^{\prime\prime}+\frac{U^{\prime}}{\eta}\right)
^{\prime}\right)  ^{\lambda}=\eta U\;.\label{radial11}%
\end{equation}
Analogously to (\ref{algo}) we impose
\begin{equation}
U(0)=1,\hspace{1cm}U^{\prime}(0)=0,\hspace{1cm}\lim_{\eta\rightarrow0^{+}%
}\frac{1}{\eta}(\eta U^{\prime})^{\prime}=-\kappa\;.\label{radial21}%
\end{equation}
After a suitable rescaling, the problem (\ref{radial11}), (\ref{radial21})
becomes
\begin{align}
z^{1+a}\left(  z^{\prime\prime}+\frac{z^{\prime}}{x}\right)  ^{\prime}  &
=\gamma x^{a}\;,\label{radial3}\\
z(0)  & =1,z^{\prime}(0)=0,\lim_{x\rightarrow0^{+}}\frac{1}{x}(xz^{\prime
})^{\prime}=-1.\label{radial4}%
\end{align}
We will denote the solutions to this problem by $z(x,\gamma)$. One can easily
check that the asymptotics near the
contact line are identical to the one dimensional case, and in particular, for the zero 
contact angle solutions
are given by Eq. (\ref{exp2})
(cf. \cite{BF}). When $\gamma=0$, there exists an explicit solution to
(\ref{radial3}), (\ref{radial4}) given by
\[
z(x,0)=1-\frac{x^{2}}{4}\;.
\]

If $\gamma<0$ then it is simple to show that $z^{\prime},z^{\prime\prime
},z^{\prime\prime\prime}$ are all negative whenever $z>0$, implying the
existence on compactly supported solutions to (\ref{radial3}), (\ref{radial4})
in this case.

\bigskip If $\gamma>0$ then one can write
\[
\left(  \frac{1}{x}(xz^{\prime})^{\prime}\right)  ^{\prime}=\gamma\frac{x^{a}%
}{z^{1+a}}\;.
\]
Noticing that
\[
\left(  \frac{1}{x}(xz^{\prime})^{\prime}\right)  ^{\prime}=8\sqrt{y}%
\frac{d^{2}}{dy^{2}}\left(  y\frac{dz}{dy}\right)  \;,
\]
where $y=x^{2}$ it is clear that function $y\frac{dz}{dy}$ is concave up. At
the origin $y\frac{dz}{dy}=0$ and $\frac{d}{dy}\left(  y\frac{dz}{dy}\right)
<0$. Hence, $\frac{dz}{dy}$ has at most one zero for $y>0$. This implies
$z(x,\gamma)$ can either decrease monotonically to zero or grow monotonically
after a minimum.

The use of implicit function theorem is analogous to the 1-D case and shows
the existence of zero contact angle solutions as well as non-zero contact
angle solutions. Since the proof follows the same lines, we omit the details.

The proof of uniqueness of the zero contact angle solution is similar to the 1-D 
case except for the fact that
the equivalent to inequality (\ref{ineq}) is now
\begin{equation}
V\left(  \frac{1}{\eta}\left(  \eta V^{\prime}\right)  ^{\prime}\right)
^{\prime}=(V^{\prime\prime\prime}+\frac{V^{\prime\prime}}{\eta}-\frac
{V^{\prime}}{\eta^{2}})V\leq0\;.\label{ineq2}%
\end{equation}

Without loss of generality we can assume that $V(0)>0$. One can easily verify
that $V^{\prime\prime}(0)=\frac{1}{2}\lim_{\eta\rightarrow0^{+}}\frac{(\eta
V^{\prime})^{\prime}}{\eta}$.

\textit{Case 1:}\textbf{\ }$\lim_{\eta\rightarrow0^{+}}\frac{(\eta V^{\prime
})^{\prime}}{\eta}<0$. Using the asymptotic expansion (\ref{exp2}) we can
conclude that there exists an interval $(\eta_{2},\eta_{1})$ where $V(\eta)$
and $V^{\prime\prime}(\eta)$ have a given sign and, moreover, $V(\eta
)V(\eta)^{\prime\prime}>0$ in that interval.

Let $\eta_{1}^{-}$ be a point of the interval sufficiently close to $\eta_{1}%
$. Let $\eta^{\ast}$ be the inflection point closest to $\eta_{1}$. Given the
regularity of $V$, since $V^{\prime\prime}(0)<0$, $V(0)>0$ and $V(\eta
)V(\eta)^{\prime\prime}>0$ for $\eta\in(\eta_{2},\eta_{1})$ such a point must
exist and $V^{\prime\prime}(\eta^{\ast})=0$.

We will show that $V(\eta^{\ast})V(\eta_{1}^{-})>0$ for $\eta^{\ast}<\eta
_{1}^{-}$. The proof applies to the case in which $V(\eta_{1}^{-})>0$ (the
proof in the case $V(\eta_{1}^{-})<0$ is analogous). Suppose that
$V(\eta^{\ast})<0$. This implies the existence of some point $\eta^{\ast\ast
}\in(\eta^{\ast},\eta_{1})$ at which a local maximum of $V$ is achieved and
$V^{\prime\prime}(\eta^{\ast\ast})<0$. Hence there must exist and inflection
point in $(\eta^{\ast\ast},\eta_{1})$ which contradicts the fact that
$\eta^{\ast}$ is the inflection point closest to $\eta_{1}$.

Since $V^{\prime\prime}(\eta^{\ast})$ is negative for $\eta\in(\eta^{\ast
}-\varepsilon,\eta^{\ast})$ and positive for $\eta\in(\eta^{\ast},\eta^{\ast
}+\varepsilon)$ we have that $VV^{\prime\prime}$ is increasing at $\eta^{\ast
}$ and hence $\left.  \frac{d}{d\eta}(VV^{\prime\prime})\right|  _{\eta^{\ast
}}=V(\eta^{\ast})V^{\prime\prime\prime}(\eta^{\ast})+V^{\prime}(\eta^{\ast
})V^{\prime\prime}(\eta^{\ast})=V(\eta^{\ast})V^{\prime\prime\prime}%
(\eta^{\ast})\geq0$ and $V^{\prime}(\eta^{\ast})V(\eta^{\ast})<0$. This
implies $(V^{\prime\prime\prime}(\eta^{\ast})+\frac{V^{\prime\prime}%
(\eta^{\ast})}{\eta^{\ast}}-\frac{V^{\prime}(\eta^{\ast})}{\eta^{\ast}{}^{2}%
})V(\eta^{\ast})>0$ which contradicts (\ref{ineq2}).

\textit{Case 2}:\textbf{\ }$\lim_{\eta\rightarrow0^{+}}\frac{(\eta V^{\prime
})^{\prime}}{\eta}>0.$ In the interval $(0,\eta_{1})$ there must exists a
local maximum $\eta_{2}$ for $V$ and we can repeat the argument of the
previous case looking for an inflection point at the left of $\eta_{2}$
instead of $\eta_{1}$.

\section{Conclusions}

In this paper we have established the existence of solutions representing the
spreading of drops in a model for the capillary spreading of Non-Newtonian
fluids of power-law rheology. The solutions depend on the rheology exponent
$\lambda$. Here we prove that when $\lambda>1$ then for a given mass of fluid,
there exists only one solution with zero contact angle at the horizontal
substrate and infinitely many solutions with a finite contact angle. Both the
spreading rate and the height of the drop are power laws whose respective
exponents depend on $\lambda$. For the case when $\lambda<1$ such spreading
solutions do not exist. Both results are valid for both planarly and radially
symmetric drops.

The results presented here pose a number of interesting questions:

\begin{enumerate}
\item  Does these solutions represent the intermediate asymptotics of the
spreading of a compactly supported drop?. In other words, does the solution as
$t\rightarrow\infty$ depend on the initial shape of the drop?.

\item  Is there a selection criterion between zero and non-zero contact angle
solutions?. This is perhaps an stability problem that can be recast in the
following way: which of these solutions is stable with respect to small
perturbations of the initial data?.
\end{enumerate}

\section{Appendix: the traveling wave solutions in 1-D}

In this section we study the existence and asymptotic behavior of traveling
wave solutions. These solutions are relevant, since they allow the
determination of the local behavior of moving fronts near the interface. They
are of the form
\begin{equation}
u(x,t)=f(x+ct)\;.\label{gg1}%
\end{equation}
Substituting (\ref{gg1}) in (\ref{ecu}) we get the equation
\begin{equation}
cf+f^{\lambda+2}|f_{\xi\xi\xi}|^{\lambda-1}f_{\xi\xi\xi}=K\;\label{gg2}%
\end{equation}
where $K$ is a real constant. The condition (\ref{noflux}) forces us to choose $K=0$.
Since
\[
|f_{\xi\xi\xi}|^{\lambda-1}f_{\xi\xi\xi}=-cf^{-1-\lambda}%
\]
one has that $f_{\xi\xi\xi}>0$ if $c<0$ and $f_{\xi\xi\xi}<0$ if $c>0$.
Therefore,
\[
\left|  f_{\xi\xi\xi}\right|  =\left|  c\right|  ^{\frac{1}{\lambda}%
}f^{-1-\frac{1}{\lambda}}\;.
\]
and we can remove the factor $\left|  c\right|  ^{\frac{1}{\lambda}}$ by
performing the change of variable $f\rightarrow\left|  c\right|  ^{\frac
{1}{2\lambda+1}}$ $f$ . Given the translation and $\xi\rightarrow-\xi$
reflection invariance of the equation we can assume without loss of generality
that the solutions representing fronts are defined for $\xi\geq0 $ and the
front is located at $\xi=0$. We will also assume $f_{\xi\xi\xi}<0$ and arrive
to
\begin{equation}
f_{\xi\xi\xi}=-f^{-1-\frac{1}{\lambda}}\;.\label{ecccu}%
\end{equation}
Let us define
\begin{align*}
x  & =f\\
y  & =f^{-\alpha}f_{\xi}\\
z  & =f^{\beta}f_{\xi\xi}%
\end{align*}
with $\alpha=\frac{\lambda-1}{3\lambda}$ and $\beta=\frac{\lambda+2}{3\lambda
}$ together with the change of variable defined by
\begin{equation}
x^{-(\beta+\alpha)}d\xi=d\xi_{1}\label{rollo1}%
\end{equation}
so that the solutions of (\ref{ecccu}) are orbits of the following third order
autonomous dynamical system:
\begin{align}
x^{\prime}  & =xy\label{dyn1}\\
y^{\prime}  & =z-\alpha y^{2}\label{dyn2}\\
z^{\prime}  & =-1+\beta yz\label{dyn3}%
\end{align}

Notice that equations (\ref{dyn2}), (\ref{dyn3}) are uncoupled to equation
(\ref{dyn1}) so that the problem reduces to the analysis of the phase plane
for equations (\ref{dyn2}), (\ref{dyn3}).

\begin{figure}[tbp]
\centerline{
\includegraphics[
height=2.1975in,width=2.4941in
]{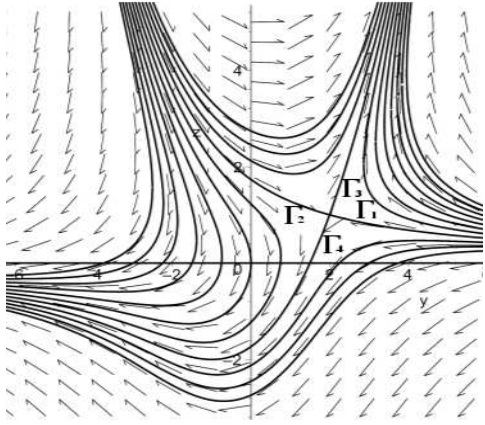}}
\caption{The phase plane for (\ref{dyn2}), (\ref{dyn3})}
\label{figura}
\end{figure}

There exists a unique equilibrium point at $P:(y,z)=(\left(  \beta
\alpha\right)  ^{-\frac{1}{3}},\alpha^{\frac{1}{3}}\beta^{-\frac{2}{3}})$.
This point is a saddle point as one can easily check. Hence, there exists an
stable manifold $\Sigma_{1}$ through $P$ as well as an unstable manifold
$\Sigma_{2}$. $\Sigma_{1}$ consists on the union of two separatrices
$\Gamma_{1}$, $\Gamma_{2}$ and $\Sigma_{2}$ on the union of another two
separatrices $\Gamma_{3}$, $\Gamma_{4}$ (see figure \ref{figura}). There exist
four different asymptotic behaviors for the trajectories as $\left|  y\right|
$ tends to infinity and each of these asymptotic behaviors corresponds to the
one exhibited by $\Gamma_{1}$, $\Gamma_{2}$, $\Gamma_{3}$, $\Gamma_{4}$
respectively. Now we proceed to describe these behaviors in detail:
%TCIMACRO{\FRAME{ftbpFU}{2.4941in}{2.1975in}{0pt}{\Qcb{The phase plane for
%(\ref{dyn2}), (\ref{dyn3})}}{\Qlb{figura}}{grf21.eps}%
%{\special{ language "Scientific Word";  type "GRAPHIC";  display "USEDEF";
%valid_file "F";  width 2.4941in;  height 2.1975in;  depth 0pt;
%original-width 5.1673in;  original-height 5.1673in;  cropleft "0";
%filename 'grf21.eps';file-properties "XNPEU";}}}%
%BeginExpansion
%EndExpansion

1) $\Gamma_{1}$ and $\Gamma_{4}$ are such that $y\rightarrow\pm\infty$ and
$z\rightarrow0^{\pm}$ respectively. Then, from (\ref{dyn2}), (\ref{dyn3}) one
obtains
\[
y^{\prime}\sim -\alpha y^{2}\;,\;z^{\prime}\sim -1+\beta yz
\]
from which it follows
\begin{equation}
\frac{dz}{dy}\sim \frac{1-\beta yz}{\alpha y^{2}}\Rightarrow z\sim
Cy^{-\frac{\beta}{\alpha}}-\frac{1}{(\alpha-\beta)y}\sim-\frac{1}%
{(\alpha-\beta)y}\;\text{as }y\rightarrow\pm\infty\label{gh1}%
\end{equation}
since $\frac{\beta}{\alpha}=\frac{\lambda+2}{\lambda-2}>1$.

2) $\Gamma_{3}$ and $\Gamma_{2}$ are such that $z\rightarrow+\infty$ and
$y\rightarrow\pm\infty$ respectively. Then, from (\ref{dyn2}), (\ref{dyn3})
one obtains
\[
z^{\prime}\sim \beta yz\;,\;y^{\prime}\sim z-\alpha y^{2}%
\]
from which it follows
\begin{equation}
\frac{dz}{dy}\sim \frac{\beta yz}{z-\alpha y^{2}}\Rightarrow(\alpha
+\frac{\beta}{2})y^{2}\sim C_{1}z^{-\frac{2\alpha}{\beta}}+z\Rightarrow
z\sim (\alpha+\frac{\beta}{2})y^{2}\;\text{as }y\rightarrow\pm\infty
\label{gh2}%
\end{equation}
since $\frac{2\alpha}{\beta}>0$.

Once we have discussed the asymptotic properties of the trajectories in the
phase plane corresponding to (\ref{dyn2}), (\ref{dyn3}), we proceed to study
the behavior of the trajectories in the three-dimensional phase space for
(\ref{dyn1}), (\ref{dyn2}), (\ref{dyn3}). Every manifold of the form
$\mathbb{R}\times\Gamma$ with $\Gamma$ being a trajectory in the phase plane,
is invariant. As we will see, the most interesting case corresponds to
$\Gamma=\Gamma_{i}\;(i=1,2,3,4)$. The invariant manifolds $\Pi_{i}%
\equiv\mathbb{R}\times\Gamma_{i}$ intersect at the points $\mathbb{R}\times P$
which form the trajectory $(x,y,z)=(e^{\left(  \beta\alpha\right)  ^{-\frac
{1}{3}}\xi_{1}},\left(  \beta\alpha\right)  ^{-\frac{1}{3}},\alpha^{\frac
{1}{3}}\beta^{-\frac{2}{3}})$. The behavior of the trajectories on $\Pi_{i}$
($i=1,2,3,4 $) are rather different. The trajectories in $\Pi_{1}$ and
$\Pi_{2}$ approach asymptotically $\mathbb{R}\times P$ as $x\rightarrow
+\infty$. From equations (\ref{dyn1}), (\ref{dyn2}) and the asymptotic
behavior computed in formula (\ref{gh1}) it follows that $y\simeq Kx^{-\alpha
}$ as $x\rightarrow0^{+}$ ($K$ an arbitrary positive constant) for the
trajectories in $\Pi_{1}$. Analogously, from equations (\ref{dyn1}),
(\ref{dyn2}) and the asymptotic behavior computed in formula (\ref{gh2}) it
follows that $y\simeq Kx^{\frac{\beta}{2}}$ when $y\rightarrow-\infty$ ($K $
an arbitrary negative constant) for the trajectories in $\Pi_{2}$. The
trajectories in $\Pi_{3}$ and $\Pi_{4}$ start at $(x,y,z)=(0,\left(
\beta\alpha\right)  ^{-\frac{1}{3}},\alpha^{\frac{1}{3}}\beta^{-\frac{2}{3}}%
)$. Analogously to the trajectories in $\Pi_{2}$ and $\Pi_{1}$, the
trajectories in $\Pi_{3}$ are such that $y\simeq Kx^{\frac{\beta}{2}}$ when
$x\rightarrow\infty$ ($K$ an arbitrary negative constant) and the trajectories
in $\Pi_{4}$ are such that $y\simeq Kx^{-\alpha}$ as $x\rightarrow0^{+}$ ($K$
an arbitrary positive constant).

Finally, we translate the phase space trajectories described above into
solutions of (\ref{ecccu}) and discuss their physical significance. The
trajectory in $\mathbb{R}\times P$ is such that, by (\ref{rollo1}) and
(\ref{dyn1}), we have
\[
\frac{dx}{d\xi}=x^{1-(\beta+\alpha)}\left(  \beta\alpha\right)  ^{-\frac{1}%
{3}}%
\]
which implies, imposing $x(0)=0$, that
\[
f(\xi)=x(\xi)=\left[  \left(  \alpha+\beta\right)  \left(  \beta\alpha\right)
^{-\frac{1}{3}}\right]  ^{\frac{1}{\left(  \alpha+\beta\right)  }}\xi
^{\frac{1}{\alpha+\beta}}\equiv C_{\lambda}\xi^{\frac{3\lambda}{2\lambda+1}%
}\;.
\]

The trajectories in $\Pi_{1}$ are such that $f(\xi)\sim C_{\lambda}\xi
^{\frac{3\lambda}{2\lambda+1}}$ as $\xi\rightarrow\infty$ and, given that
$y\simeq Kx^{-\alpha}$ as $x\rightarrow0^{+}$, by (\ref{rollo1}) and
(\ref{dyn1}) one has
\[
\frac{dx}{d\xi}=K
\]
so that $f(\xi)\sim K\xi$ as $\xi\rightarrow0^{+}$.

The trajectories in $\Pi_{2}$ are such that $f(\xi)\sim C_{\lambda}\xi
^{\frac{3\lambda}{2\lambda+1}}$ as $\xi\rightarrow+\infty$ and $f(\xi)\sim
K\xi^{2}$ as $\xi\rightarrow-\infty$. Since there is no front in this case, we
do not consider this solution to be physically relevant.

The trajectories in $\Pi_{3}$ are such that $f(\xi)\sim C_{\lambda}\xi
^{\frac{3\lambda}{2\lambda+1}}$ as $\xi\rightarrow0^{+}$ and $f(\xi)\sim
K\xi^{2}$ as $\xi\rightarrow\infty$.

The trajectories in $\Pi_{4}$ are such that $f(\xi)\sim C_{\lambda}\xi
^{\frac{3\lambda}{2\lambda+1}}$ as $\xi\rightarrow0^{+}$ and $f(\xi)\sim
K(\xi_{0}-\xi)$ as $\xi\rightarrow\xi_{0}^{-}$ for some positive $\xi_{0}$.
They are compactly supported.

In addition there exist another four families of solutions such that the
trajectories in phase space approach asymptotically to two of the manifolds
$\Pi_{i}$ ($i=1,2,3,4$). The first one approaches $\Pi_{1}$ and $\Pi_{4}$ so
that behaves linearly at the origin and linearly close to some $\xi_{0}>0 $.
It is compactly supported and lacks a clear physical significance. The second
one approaches $\Pi_{4}$ and $\Pi_{2}$ so that it is linear at the origin and
grows quadratically at infinity. For them $y<0$ which implies that $f^{\prime}%
(\xi)<0$. They represent dewetting solutions. The third one approaches
$\Pi_{2}$ and $\Pi_{3}$ and grows quadratically at $\pm\infty$ presenting no
fronts. The last one approaches $\Pi_{3}$ and $\Pi_{1}$. It behaves linearly
at the origin and grows quadratically at infinity.

To summarize, one has infinitely many solutions that behave linearly at the
origin and such that $f(\xi)\sim C_{\lambda}\xi^{\frac{3\lambda}{2\lambda+1}}$
as $\xi\rightarrow\infty$. There is only one solution (which is explicit) such
that $f(\xi)\sim C_{\lambda}\xi^{\frac{3\lambda}{2\lambda+1}}$ at the origin
(zero contact angle) and at infinity. There exist infinitely many solutions
with zero contact angle at the origin and growing quadratically at infinity
and, finally there exist dewetting solutions which are linear at the origin
and grow quadratically at infinity. Hence, the only local behaviors near the
interface for moving fronts are those with finite contact angle and the one
with zero contact angle.\\

{\bf ACKNOWLEDGMENTS} We thank the IMA (University of Minnesota), the Mathematics Department of the
University of North Texas
and the Departament of Applied Mathematics of Universidad Rey Juan Carlos for their support and 
use of their facilities.\\

\end{document}